\documentclass[useAMS,usenatbib,usegraphicx]{mn2e}
\usepackage{times}
\usepackage{amsmath,amssymb}

\begin{document}
 \title[Locating the VHE source in the Galactic Centre]{Locating the VHE source in the Galactic Centre 
        with milli-arcsecond accuracy}
 \author[Abramowski et al.]{A.~Abramowski$^1$, S.~Gillessen$^2$, D.~Horns$^1$, H.-S.~Zechlin$^1$\\
$^1$Department of Physics, University of Hamburg, Luruper Chaussee 149, D-22761 Hamburg, Germany \\
$^2$MPI for Extraterrestrial Physics, Giessenbachstrasse, D-85748 Garching, Germany }
 \maketitle
 \begin{abstract}
Very high-energy $\gamma$-rays (VHE; E$>$100 GeV) have been detected from the
direction of the Galactic Centre up to energies E$>$10 TeV. Up to now, the
origin of this emission is unknown due to the limited positional
accuracy of the observing instruments. One of the counterpart candidates is the
super-massive black hole (SMBH) \mbox{Sgr A$^{*}$}.   
If the VHE emission is produced within $\approx10^{15}$ cm $\approx1000 \ r_G$ 
($r_G=G M/c^2$ is the Schwarzschild radius) of the SMBH, a decrease of the
VHE photon flux in the energy range 100--300 GeV is expected whenever
an early type or giant star approaches the line of sight within $\approx$ milli-arcseconds (mas).
The dimming of the flux is due to absorption by pair-production of the VHE photons 
in the soft photon field of the star, an effect we refer to as pair-production eclipse (PPE).
Based upon the currently known orbits of stars in the inner arcsecond of the Galaxy we find that PPEs 
lead to a systematic dimming in the 100--300 GeV band at the level of a
few per cent and lasts for several weeks.  Since the PPE affects only a narrow energy band
and is well correlated with the passage of the star, it can be clearly discriminated against other 
systematic or even source-intrinsic effects. 
While the effect is too small to be observable with the current generation of VHE detectors, upcoming high
count-rate experiments like the Cherenkov telescope array (CTA) will be
sufficiently sensitive. Measuring the temporal signature of the PPE bears the
potential to locate the position and size of the VHE emitting region within the
inner 1000 $r_G$ or in the case of a non-detection exclude the immediate environment of the SMBH
as the site of $\gamma$-ray production altogether.  
\end{abstract}


\section{Introduction}
The central region of our Galaxy harbours a super-massive black hole (SMBH) with a mass 
\mbox{$M=(4.31\pm 0.06_{|\,\mathrm{stat}}\pm0.36_{\,|R_0})\times 10^{6} M_\odot$}
at a distance from Earth of $R_0=8.33\pm 0.35$ kpc (\citealt{2009ApJ...692.1075G}).
Its proximity enables  detailed observations with
high angular resolution in the radio and, recently by the use of adaptive
optics, in the near infrared band (for a review see e.g. \citealt{2009IJMPD..18..889R}). 
Accurate measurement of the Keplerian orbits
of stars in the inner arcsecond of the Galactic Centre
 \citep{2002Natur.419..694S,2003ApJ...586L.127G,Eisenhauer2005,2009ApJ...692.1075G}
has revealed the presence of a compact massive object co-located with the
radio-source \mbox{Sgr A$^{*}$}
\citep{1974ApJ...194..265B,2007ApJ...659..378R}.  The astrometry allows for a
shift of up to 2 mas \citep{2009ApJ...692.1075G} being the upper limit on the
systematic error of the position between the radio and near infrared positions
of \mbox{Sgr A$^{*}$}. 

\begin{figure}
\begin{center}
  \includegraphics[width=80mm,bb=0 53 695 580, clip=]{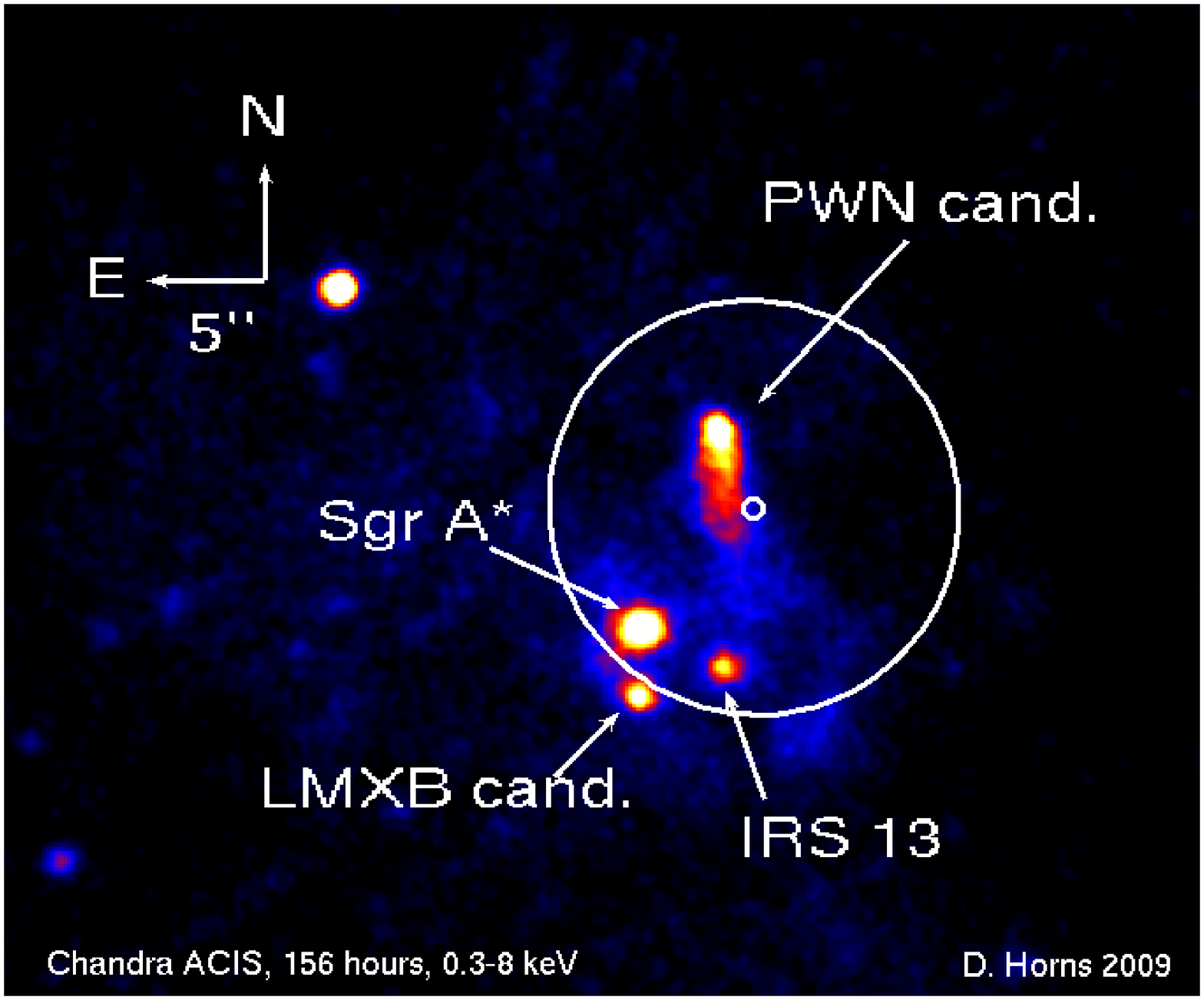}
  \caption{\textit{Chandra} X-ray image of the Galactic Centre region accumulated from 156 hours of data in the 0.3--8 keV band. 
The possible sources for the VHE emission are marked, the small circle 
indicates the position of the H.E.S.S. source
 (\mbox{$l=359^\circ 56^\prime 41.1^{\prime\prime} 
\pm 6.4^{\prime\prime}$ (stat.)},
 \mbox{$b=-0^\circ 2^\prime 39.2^{\prime\prime} 
\pm 5.9^{\prime\prime}$ (stat.)}) as reported in \citet{GCPosition}. 
The large white ellipse marks the combined statistical and systematical (6 arcsec) 
H.E.S.S. positional uncertainty region. The supernova candidate Sgr A East is outside of the image.}
  \label{fig:chandra}
\end{center}

\end{figure}

Recently, very high-energy (VHE; E$>$100 GeV) emission  has been detected from the Galactic Centre region
\citep{2004A&A...425L..13A,2004ApJ...608L..97K,2004ApJ...606L.115T,2006ApJ...638L.101A}.
A number of different scenarios for particle acceleration and \mbox{$\gamma$-ray} production have been discussed in the literature.
\citet{2005ApJ...619..306A} have considered \mbox{$\gamma$-ray} production
both from electrons via curvature radiation as well as from hadronic processes 
(photo-meson production and proton-proton scattering). The energetic photons
can be produced as close as 20 $r_G$ without suffering strong absorption via pair-production with
the photon field generated by the low-luminosity SMBH ($L<10^{-8}~L_\mathrm{Edd}$) \citep{2005ApJ...619..306A}.
In another SMBH-related scenario suggested by \citet{2004ApJ...617L.123A} an assumed relativistic
wind injected in the vicinity of the SMBH would lead to the formation of a synchrotron/inverse-Compton
nebula similar to pulsar wind generated nebulae. In this case the VHE emission would be produced
in the vicinity of the shock that is leading to acceleration of particles at a distance of $\mathcal{O}(10^{16})$ cm
to the SMBH.
More production scenarios, which are not directly related to the SMBH, have been discussed:
The pulsar-wind nebula candidate G359.95-0.04 \citep{2006MNRAS.367..937W}, the stellar cluster IRS 13
\citep{2004A&A...423..155M}, the low-mass X-ray binary candidate
J174540.0-290031, Sgr A East \citep{2006PhRvL..97v1102A}
or self-annihilating Dark Matter (see e.g. \citealt{2005PhLB..607..225H}).
\mbox{Fig. \ref{fig:chandra}} shows an X-ray image (0.3--8~keV)  
derived from 156~hours of archival \textit{Chandra} observations including the \mbox{X-ray} emission observed from
the various counterpart candidates located within the location uncertainty of the VHE source.

Recent improved reconstruction of the location of the VHE-source 
\citep{GCPosition} has excluded the emission originating
predominantly from Sgr A East, while the other possibilities mentioned above
remain viable.

The absence of flux variability in the VHE  band during simultaneous observations of Sgr A$^*$ by H.E.S.S. (High Energy Stereoscopic Systme) and \textit{Chandra} seem to disfavour a common origin of the X-ray and the VHE emission, yet such a scenario can not
be excluded \citep{2008A&A...492L..25A}.

Despite the efforts to reduce the
positional error of the VHE source, the current uncertainty of 6 arcsec (syst.) per 
telescope axis \citep{GCPosition} is limited to systematic errors. 
The prospects for a substantial improvement on the systematic uncertainties are limited by
the mechanical stability of air Cherenkov telescopes, which would require substantial improvement
with respect to the currently used light-weight construction.
The Galactic Centre region is too crowded to identify the 
VHE source reliably. Unless a clear timing signature can be established, this will not change.

In this paper we propose the use of a time-dependent attenuation feature, which provides a novel opportunity
to constrain the size and location of the Galactic Centre VHE-source with an accuracy of milli-arcseconds
in the case of the VHE emission being produced in the vicinity of Sgr A$^*$. 
We provide a long-term (60 yrs) prediction of flux variations in different energy bands which can be caused
by the passage of stars close to the line of sight towards the VHE source.

\section{Pair-production eclipses}
In the following we assume that VHE $\gamma$-rays originate from a region near the SMBH \mbox{Sgr A$^{*}$}. 
These VHE photons interact with soft photon fields via the pair-production process
($\gamma_\mathrm{{VHE}}+\gamma_{*}\rightarrow e^{+}+e^{-}$). Stars on stable
Keplerian orbits in the direct vicinity of \mbox{Sgr A$^{*}$}
($<1^{\prime\prime}$) provide these (low-energy) photon fields, and
pair-production can lead to an  energy-dependent dimming of the
VHE-flux.  Since the position of the star changes, the
optical depth $\tau$ for the VHE photon varies along the orbit of the star.
In the following we will calculate $\tau$ as a function of time and $\gamma$-ray energy for stars
on Keplerian orbits.
\subsection{Pair-production}
\begin{figure}
\includegraphics[width=0.95\linewidth]{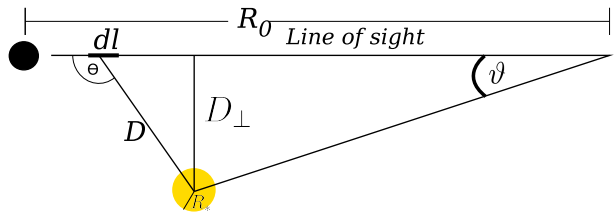}
\caption{\label{fig:sketch} A cartoon of the line of sight integration and the used geometry.}
\end{figure}
The total cross-section for pair-production (after averaging over polarisation states) is given as
\begin{equation}
\begin{split}
 \sigma(\varepsilon,E,\theta)&=\frac{3\sigma_T}{16}(1-\beta^2) \\ &\times \left[(3-\beta^4)\ln\left(\frac{1+\beta}{1-\beta}\right) \
-2\beta(2-\beta^2)\right],
\end{split}
\end{equation}
with $\sigma_T$ being the Thomson cross-section, $\beta=(1-1/s)^{1/2}$ and
$s=\varepsilon E/(2m_e^2c^4) (1-\cos \theta)$ the (normalized) centre-of-momentum energy of two photons
of energy $\varepsilon$ and $E$ interacting at an angle $\theta$ in the
laboratory frame \citep{1967PhRv..155.1404G}.\footnote{In this paper
$\varepsilon$ denotes the energy of the low-energy stellar photon, $E$ is the
energy of the high-energy photon coming from \mbox{Sgr A$^{*}$}; $m_e$ is the electron mass.} We choose the
laboratory frame as the system where the VHE source is at rest. The
cross-section is strongly energy-dependent and  has a maximum for photon
energies of
\begin{equation}
\label{eqn:emax}
  E_{\mathrm{max}}=0.9 \left( \frac{\varepsilon}{\mathrm{eV}}\right)^{-1}  \mathrm{TeV,}
\end{equation}
corresponding to a maximal cross-section for a 1 TeV $\gamma$-photon
interacting with a 0.9 eV ($\lambda\approx 1.38 \ \mathrm{\mu m}$, near
infrared) stellar photon. 
The relative velocities of the observer  as well as  the star with respect to the VHE source are considered to be
small (with respect to the speed of light $c$) and will be neglected in the following.  For the
considered case we assume  the stellar atmosphere of a star with radius $R_\ast$
and effective temperature $T_\mathrm{eff}$ to be the source
of low-energy photons. The differential photon density
$dn=n(\varepsilon)~d\varepsilon$ at the photosphere of the star can
 be approximated with a black-body spectrum:
\begin{equation}
 n(\varepsilon)=\frac{2\varepsilon^{2}}{h^{3}c^{3}}\left[\exp\left({\frac{\varepsilon}{k_BT_{\text{eff}}}}\right)-1\right]^{-1},
\label{eqn:bbody}
\end{equation}
with $h$ being Planck's constant, $c$ the speed of light and $k_B$ 
Boltzmann's constant. The optical depth for VHE photons with energy $E$ 
is calculated by integrating 
along the line of sight subtending a perpendicular distance $D_\perp$ to the star (see Fig. \ref{fig:sketch} for
a definition of the geometry):
\begin{equation}
\tau(E)=\int\limits_{0}^{R_0} dl   \int\limits_{c_{\text{min}}}^{1} d\cos{\Theta} 
\int\limits_{0}^{2\pi} d\phi  \int\limits_{\varepsilon_{\text{min}}}^{\infty} d
\varepsilon \ \sigma(\varepsilon,E,\theta) \ n(\varepsilon) \ (1-
\cos{\theta}), \label{eq:tau}
\end{equation}
with $c_\mathrm{min}=\left[1-\left(\frac{R_\ast}{D}\right)^2\right]^{1/2}$ 
and $\varepsilon_\mathrm{min}=2m_e^2c^4/[E(1-\cos{\theta})]$ being the 
minimum energy for pair-production to occur \citep[e.g.][]{2006A&A...451....9D}. 
For $R_\ast\ll D$ the integration over the solid angle in Eq. \ref{eq:tau} leads to the point-source approximation:
\begin{equation}
\tau(E)=\int\limits_{0}^{R_0} dl  \int\limits_{\varepsilon_{\mathrm{min}}}^{\infty}  d
\varepsilon \ \pi \left(\frac{R_\ast}{D}\right)^2   \ \sigma(\varepsilon,E,\theta) \ n(\varepsilon) \ (1-\cos{\theta}). 
\label{eq:tau_point}
\end{equation}
For the distances examined here this
is an appropriate approximation and will be used in the following.
\begin{figure}
\includegraphics[width=0.95\linewidth]{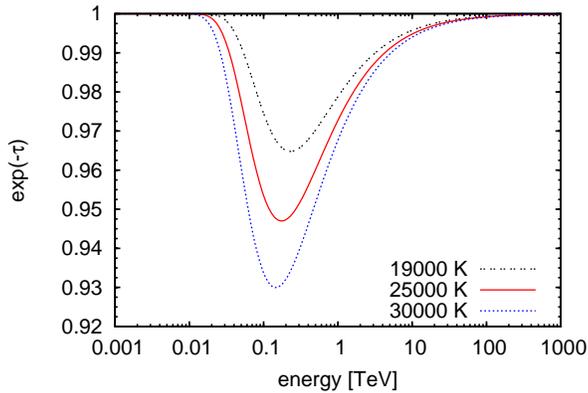}
\caption{\label{fig:tautemp} Energy dependence of the absorption
$\exp{(-\tau)}$ for a fixed perpendicular angular separation of 1 mas (for a distance
$R_0=8.33$ kpc). The three curves show the absorption for 
temperatures $T_{\mathrm{eff}}$ of the star; the stellar radius is adjusted
according to the range of stellar types suggested for the well-studied S2 star
 ($T_{\mathrm{eff}}=19\,000$ K, $R_\ast=13.1~R_\odot$;
$T_{\mathrm{eff}}=25\,000$ K, $R_\ast=10.7~R_\odot$; $T_{\mathrm{eff}}=30\,000$ K,
$R_\ast=9.4~R_\odot$) \citep{2008ApJ...672L.119M}.}
\end{figure}
In order to demonstrate the uncertainty on the optical depth that results from the uncertainty
on the stellar parameters, we calculate the optical depth for the well-studied star ``S2''. This
star is constrained to have a temperature $19\,000 \ \mathrm{K}<T_{\mathrm{eff}}<30\,000 \ \mathrm{K}$ and therefore stellar radii $13.1 \ R_\odot>R_\ast>9.4 \ R_\odot$
\citep{2008ApJ...672L.119M} corresponding to spectral types B0--B2.5V. 
Fig. \ref{fig:tautemp} shows the result of the calculation of
Eq.~\ref{eq:tau_point} for a fixed distance $D_\perp $ corresponding to a projected angular
separation of 1 mas of the VHE source and the star. 
The absorption shows a pronounced minimum around 200 GeV for $T_{\mathrm{eff}}=19\,000$ K shifting to smaller energies 
(see also Eq. \ref{eqn:emax}) with increasing stellar temperature. The stellar types have been determined spectroscopically for nearly 50 so-called S-stars.
Most stars are early type,  B-main-sequence-stars with stellar radii of
10--20 $R_\odot$ and effective temperatures of 2--3$\times 10^4$ K
(\citealt{2008ApJ...672L.119M}).

As shown in Appendix \ref{AppA} an additional contribution of cascade radiation coming from the electron/positron pairs produced in the PPE is negligible.
The resulting change is similar to the uncertainties
resulting from the uncertainties of the orbital elements. Therefore, the
effect is generally not of importance.

Another effect which could change the timing signature is a possible change of the accretion flow towards the SMBH induced by the star closely passing \mbox{Sgr A$^{*}$}. Direct wind-driven accretion can lead to an increased accretion rate, but the fast motion of the S-stars does not allow any accretion during periastron passage \citep{2004MNRAS.350..725L}. Furthermore, the PPE can be treated as independent from accretion induced variability, because it is strongly time- and energy-dependent.

\subsection{Stellar orbits and optical depth for VHE photons}
The temporal evolution of the optical depth calculated by means of Eq. \ref{eq:tau_point} depends on the
position of the star relative to the line of sight, which in turn varies along
the orbit of the star. The orbital parameters of 29 stars in the
central arcsecond of the Milky Way have been well-measured  
using both spectroscopy and time-dependent astrometry with unprecedented accuracy
utilising the VLT advanced adaptive optics in the near-infrared (\textit{K}-)band 
\citep{Eisenhauer2005,2009ApJ...692.1075G}. 

\begin{figure}
\includegraphics[width=0.95\linewidth]{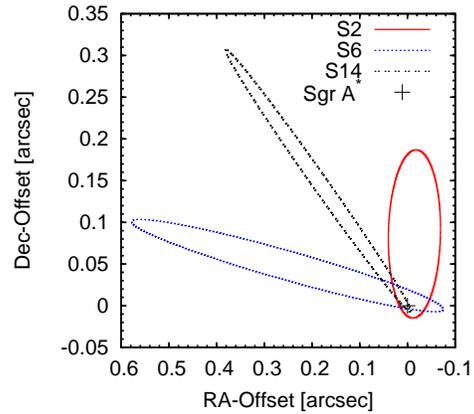}
\caption{\label{fig:orbit} The projected (on the plane of the sky) orbits of three selected
S-stars (see text). The cross marks the position of \mbox{Sgr
A$^{*}$}.}
\end{figure}

With the given orbital parameters we derived the three-dimensional position using the known methods described e.g. in \citet{Murray}. The projection of three stellar orbits on the plane of the sky is shown in
Fig.~\ref{fig:orbit}.  The stars S2, S6, and  S14 have been selected in particular, because they
approach the line of sight closest and therefore are expected to lead to the
largest absorption effect.  Since the spectral types of these stars are not exactly known,
representative values for a B-type main sequence star (similar to the parameters for S2) are used: 
$R_\ast=10.7~R_\odot$ and $T_{\mathrm{eff}}=2.5 \times 10^4$ K.

\begin{table*}
\begin{minipage}{180mm}
\begin{scriptsize}
\caption{Parameters of selected S-stars: The first 7 columns show the 
\label{tab:parameter}
orbital parameters taken from \citet{2009ApJ...692.1075G}, 
the last two columns list the minimal angular separation of the star to the line of sight $\vartheta$ (in mas)
and the resulting maximum absorption $\exp{[-\tau(E)]}$ for $E=200~\mathrm{GeV}$.}
\begin{center}
\begin{tabular}{|l||c|c|c|c|c|c|c|c|c|}
\hline
Star & $a~[^{\prime\prime}]$ & $e$ & $i~[^{\circ}]$ &
$\Omega~[^{\circ}]$ &$\omega~[^{\circ}]$& $t_P~[yr]$ & $T~[yr]$ & $\vartheta$[mas] & $\exp{(-\tau)}$\\
\hline
S2  & $0.123 \pm 0.001$ & $0.880 \pm 0.003$ &  $135.25 \pm 0.47$ & $225.39 \pm 0.84$ & $63.56 \pm 0.84$ & $2002.32 \pm 0.01$ & $15.80 \pm0.11$ & $10.8\pm0.4$& $0.9944^{+0.0003}_{-0.0002}$ \\
S6  & $0.436 \pm 0.153$ & $0.886 \pm 0.026$ &  $86.44 \pm 0.59$  &
$83.46 \pm 0.69$  & $129.5 \pm 3.1$ & $2063 \pm 21$ & $105 \pm 34$ & $3.4\pm0.9$  & $0.966^{+0.002}_{-0.008}$\\
S14 & $0.256 \pm 0.010$ & $0.963 \pm 0.006$ &  $99.4 \pm 1.0$    &
$227.74 \pm 0.70$ & $339.0 \pm 1.6$ & $2000.07 \pm 0.06$ & $47.3 \pm 2.9$ & $2.2\pm0.4$ & $0.976^{+0.002}_{-0.003}$\\
\hline
\end{tabular}
\end{center}
\end{scriptsize}
\end{minipage}
\end{table*}
\begin{figure} 
\includegraphics[width=0.95\linewidth]{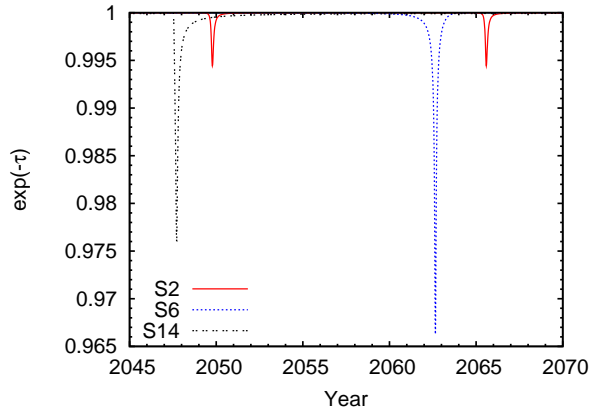}
\caption{\label{fig:absorption} Absorption $\exp[-\tau(E,t)]$ due to
pair-production eclipses for three S-stars for VHE photons of energy $E=200$ GeV as function of the calendar year.}
 \end{figure}
\begin{figure}
\includegraphics[width=0.95\linewidth]{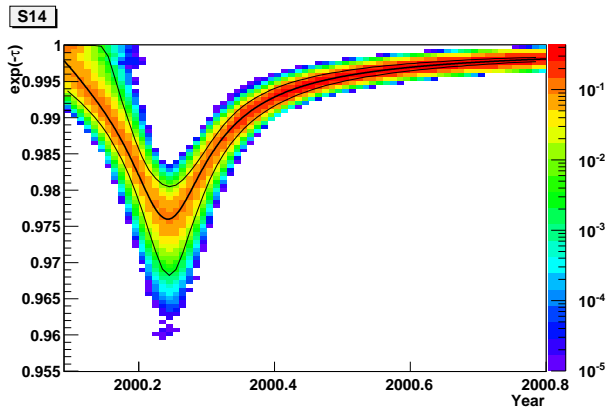}
\caption{\label{fig:s14_shifted}
Absorption $\exp[-\tau(E,t)]$ due to pair-production eclipses for the star S14: the curves denote the 
most likely absorption and the 99 per cent c.l. region. The underlying histogramme indicates the smoothed probability density
function resulting from the Monte--Carlo calculations (see text for details).}
\end{figure}
Table \ref{tab:parameter} lists the known orbital parameters,
their uncertainties, the derived minimal angular separation $\vartheta\approx D_\perp/R_0$ as well
as the correspondingly calculated absorption $\exp(-\tau)$ for VHE photons of energy $E=200$ GeV. 
The optical depth takes into account the full three-dimensional geometry of the problem. For the three stars
considered here the calculated maximum attenuation of the VHE-flux turns out to be 
0.5 per cent for S2, 2.4 per cent for S14 and 3.4 per cent for S6.
Fig. \ref{fig:absorption} depicts the predicted absorption of these three stars 
until the year 2070.

In order to propagate the uncertainties on the orbital parameters into uncertainties on the absorption, a dedicated
Monte--Carlo-type simulation has been performed. Since the errors on the different orbital parameters are correlated, we transformed the parameters
in the system of the eigenvectors of the covariance matrix and varied these transformed
(then uncorrelated) parameters randomly 10\,000 times following a normal distribution within the known variance. After
back-transforming, we end up with 10\,000 orbital parameter-sets tracing the probability density function 
including the correlation of the parameters.

In order to derive the resulting uncertainty on the absorption effect,
 we calculated $\exp[-\tau(E,t)]$ for each
parameter-set obtained in the Monte--Carlo simulation. As a consequence of the
uncertainties of the orbital parameters the time when the minimum of $\exp(-\tau)$ occurs 
varies considerably. Since we are mainly interested in the uncertainty on the absorption at a fixed time 
we shifted all curves in time to the mean (best-fitting) value.   The most likely
attenuation $\exp(-\tau)$ for the star S14 is shown in Fig. \ref{fig:s14_shifted} together with the 
99 per cent confidence interval. 
For the other stars the uncertainties (68 per cent confidence level) on the minimum 
of $\exp(-\tau)$ are listed in Table \ref{tab:parameter}.
Moreover, the often unknown exact stellar type leads to additional uncertainties on the absorption, which are of similar magnitude as the uncertainties due to the errors on the orbital elements.

\subsection{Simulated light curves}

The predicted flux modulation due to PPE is at the
level of a few per cent (depending on the estimated uncertainties of the orbit and
stellar type). In order to investigate the possibility of discovering PPEs measurements of light curves are simulated for future Imaging Air
Cherenkov Telescopes (IACTs). For completeness and comparison 
light curves for the H.E.S.S. telescope system \citep{2004NewAR..48..331H} are simulated as well.

The expected photon rate $R(E_1,E_2,t)$ in an energy band \mbox{$E_1$--$E_2$}
is calculated: 
\begin{equation}
\label{eq:rate}
 R(E_1,E_2,t)=\int\limits_{E_{1}}^{E_{2}} dE \ A_{\mathrm{eff}}(E)  \phi(E) \ \exp{[-\tau(E,t)]}
\end{equation}
with $A_{\mathrm{eff}}(E)$ the effective area of the IACT, $\phi(E)$ the
differential VHE photon spectrum from the Galactic Centre, and $\tau(E,t)$ the optical depth
calculated by means of Eq. \ref{eq:tau_point}. The photon spectrum $\phi(E)$ is assumed
to follow a power-law with photon index $2.25$ \citep{2006PhRvL..97v1102A}. 
The recently published finding of an exponential cut-off in the energy spectrum at $\approx 16$ TeV
\citep{2009A&A...503..817A} changes the predicted rate between 1 and 10 TeV only marginally and is therefore
neglected here.

\begin{figure}
\includegraphics[width=0.95\linewidth]{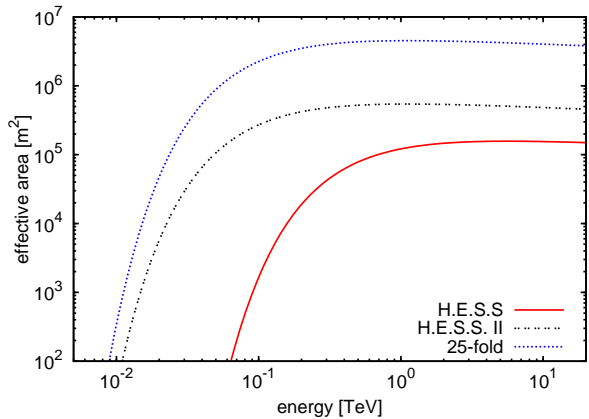}
\caption{\label{fig:area} Effective areas for different IACTs used in the simulation of the light curves.}
\end{figure}

The effective areas assumed (see Fig. \ref{fig:area}) are parameterised 
based upon \citet{2006A&A...457..899A} for H.E.S.S. and \citet{Punch2005} for
H.E.S.S. II (see Appendix \ref{AppB} for further details). For a next generation telescope-system (like the planned CTA,
see e.g. \citealt{2008ICRC....3.1313H}) we generically assumed an increase of the effective area by a factor of 25  with respect 
to the H.E.S.S. collection area as well as a lower energy
threshold of about 10 GeV.
In the following three separate energy bands are considered: 
the very low energy (10--50~GeV) band, the low energy band (100--300~GeV), and the high
energy band (1--10~TeV). Given the energy dependence of the PPE, most of the attenuation will
affect the low energy band, whereas the adjacent energy bands remain without attenuation.
\begin{figure*}
  \includegraphics[width=0.33\linewidth]{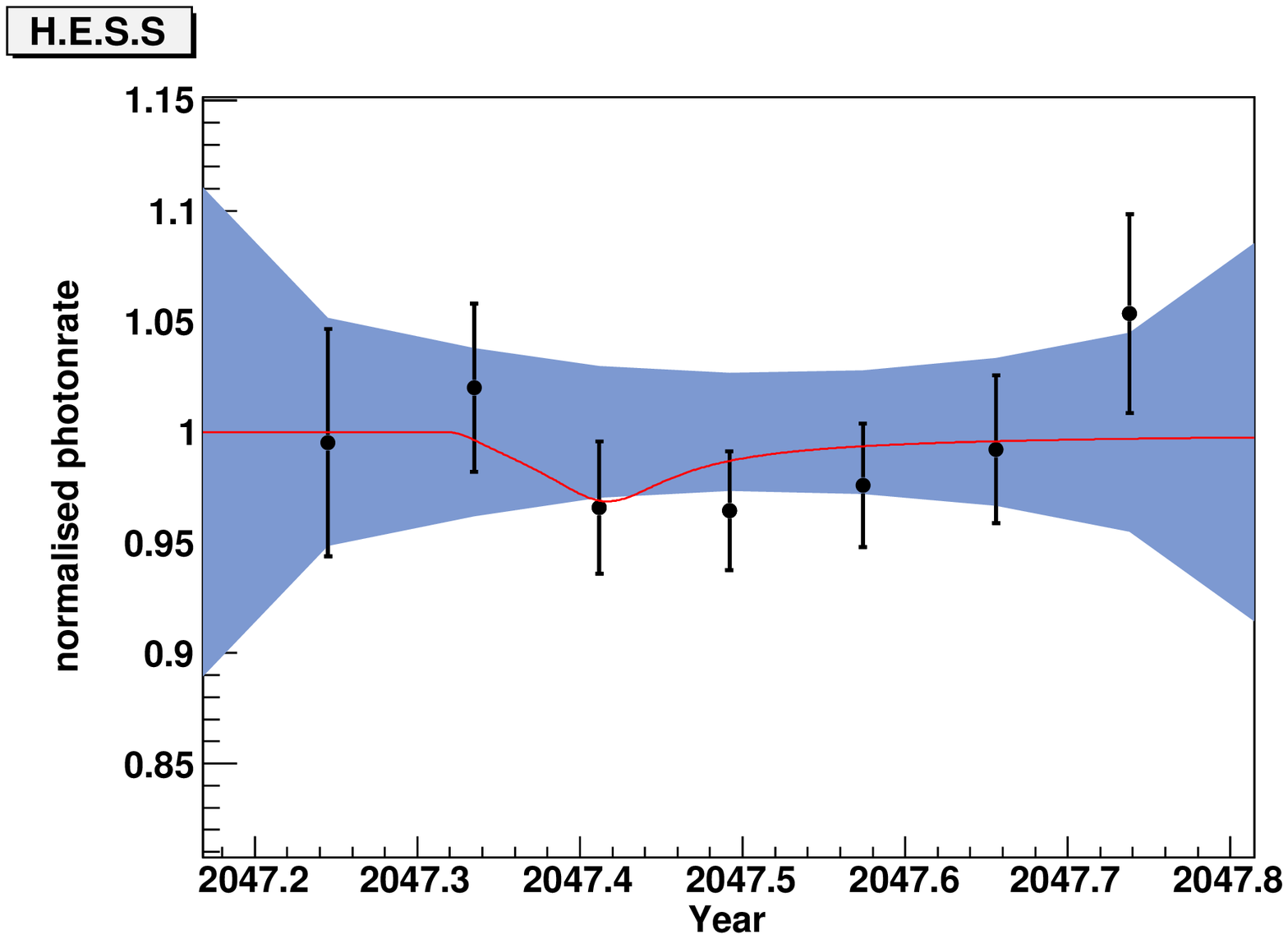} 
  \includegraphics[width=0.33\linewidth]{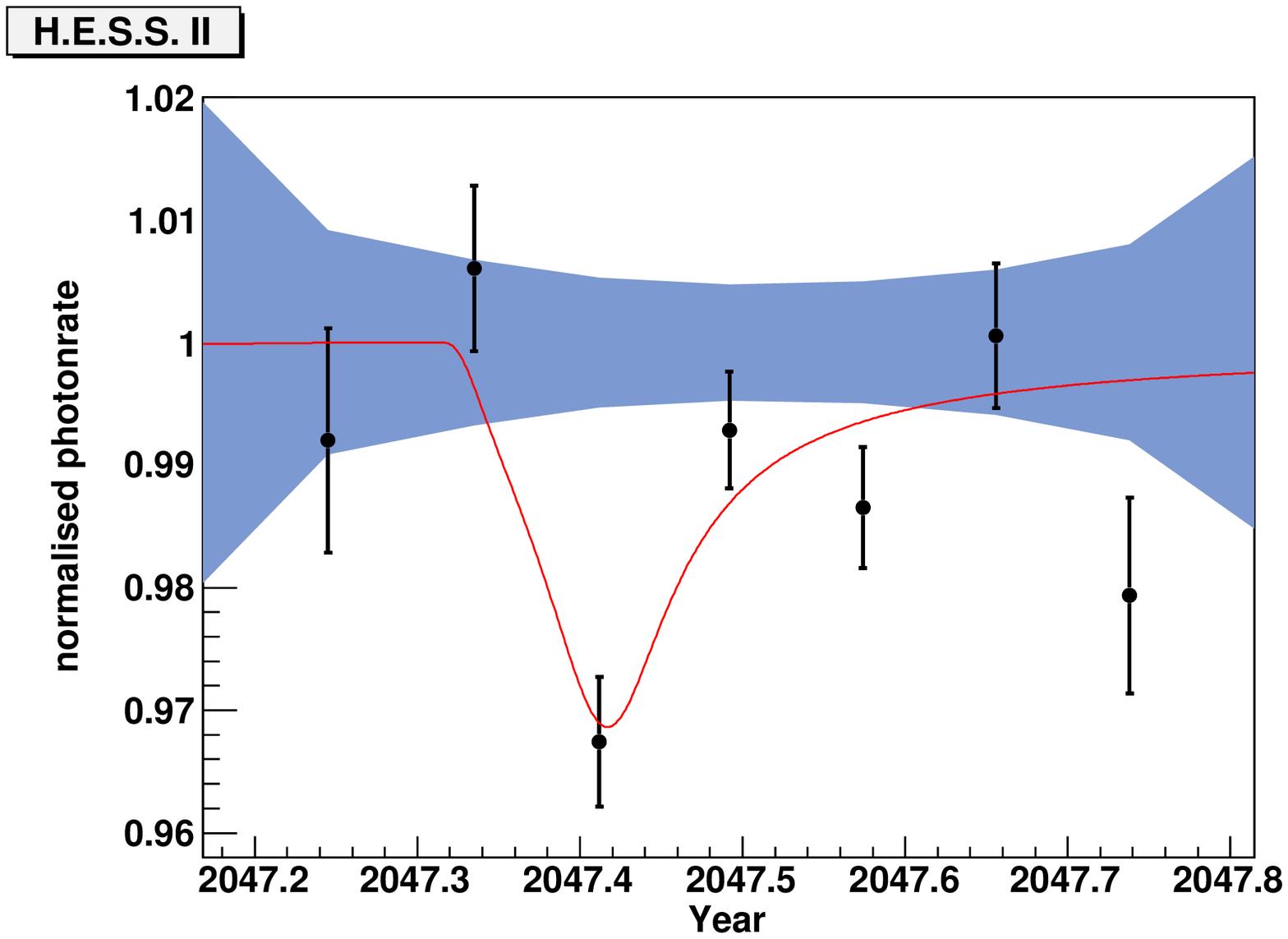} 
  \includegraphics[width=0.33\linewidth]{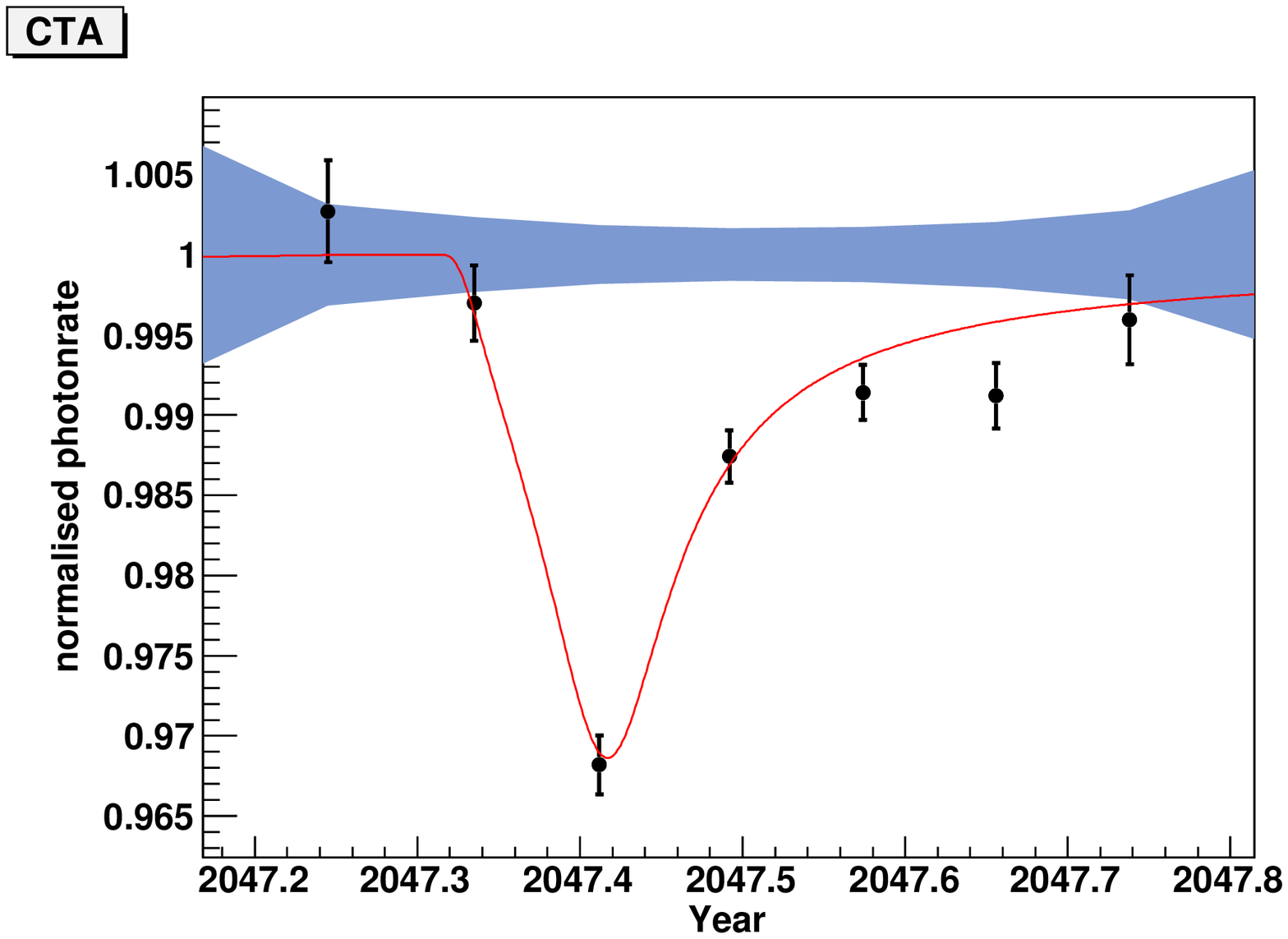} 
  \caption{Simulated light-curves for the star S14 for 3 different experiments.
The red curve shows the calculated absorption for $E=200$ GeV, the points show
the normalised photon-rate in the low energy band (100--300 GeV) where the effect
has its maximum. The shaded area marks the photon-rate and its statistical uncertainty without absorption in the
combined high (1--10 TeV) and very low (10--50 GeV) energy band. In this band the
expected absorption is negligible (see Fig. \ref{fig:tautemp}). From left to
right: H.E.S.S., H.E.S.S. II, CTA (note: change in scale of the three panels y-axis).}
  \label{fig:lightcurve}
\end{figure*}
The available observation time is calculated for a site at the Tropic of Capricorn taking into account the
dark (moon-less) visibility time  of the Galactic Centre between roughly March to 
September. For each new moon period, the number of detected photons is estimated in the three energy bands by 
calculating $N_\mathrm{ph}(E_1,E_2)=R \ t_\mathrm{vis}$, with $t_\mathrm{vis}$ being the actual observation time available
until the Galactic Centre reaches an altitude smaller than 30$^\circ$.
The variance on the number of observed photons is estimated under the assumption of Poissonian statistics, and
the expected value $N_\mathrm{ph}$ is sampled following a normal distribution.
This is done separately for the three energy bands. The resulting simulated light curve for the passage of 
S14 in the year 2047 (normalised 
to the unattenuated light curve) for the
low-energy band for the three IACT setups is shown in the three panels of Fig. \ref{fig:lightcurve}. 
In addition to the attenuated low-energy band, the other two energy bands are combined into a comparison measurement
marked by the shaded region.
This comparison measurement allows the elimination of systematic effects due to observational artefacts or
even source intrinsic, broad band variability of the source. It should be noted that the variation 
of the collection area for different zenith angles has not been included in the calculation. In principle,
for smaller altitude angles, the lowest energy band will not be accessible any more. However, this will
be compensated by an increase of the collection area at larger energies not altering the conclusions drawn here.

The large
uncertainty of the time of the closest approach to the line of sight for S14  ($\approx 3$ yrs) leaves us freedom
to shift the light-curve in time. The time of the closest approach is shifted slightly with respect to the best-fit 
value by 2 months to
the middle of the year 2047. In this way the maximum observation time available for Cherenkov telescopes  is in phase 
with the time of the closest approach.   Additionally, we assumed the largest possible
attenuation of the flux within the orbital uncertainties (see
Table  \ref{tab:parameter}). 
Fig. \ref{fig:lightcurve} shows that a state of the art IACT like H.E.S.S. (left graph) is not
able to detect a PPE within the measurement errors. Under the favourable assumptions used H.E.S.S. II (middle graph) will be able to discover
a PPE within the errors. A next generation IACT-system like
the planned CTA (right graph) will discover PPEs
even under unfavourable conditions.

 \section{Summary and discussion}
Under the assumption that a compact emission region ($<~1000~r_G$) 
in the vicinity of \mbox{Sgr A$^{*}$} is the dominant source of the $\gamma$-rays, 
pair-production eclipses can lead to a time- and energy-dependent dimming of the
$\gamma$-ray flux from the direction of the Galactic Centre.
Here, we have calculated for the first time the
effect of PPEs for the Galactic Centre and provide predictions 
for the next few decades. 

For the currently known orbits of S-stars the effect will be detectable with future IACTs. 
The next close approach to the
line of sight causing a dimming $>$ 1 per cent will be around the year 2047. 
 At the moment, S-stars to a limiting magnitude of
$m_K=18$ (\textit{K}-band) are tracked within $2^{\prime\prime}$ around
\mbox{Sgr A$^{*}$} with the VLT \citep{2009ApJ...692.1075G}, resulting in a
total number of 109 stars routinely followed. It is therefore expected that 
the orbital parameters of these, and even fainter stars, will be determined in the future;
some of these stars may lead to even stronger PPEs at an earlier time. 

So far, we have neither considered the effect of binary systems nor of  
non main sequence (e.g. giant) stars leading to a possibly larger absorption effect. 

A detection of a PPE has the potential to resolve the current source confusion at the Galactic
Centre. It could constrain the $\gamma$-ray emitting region to the direct
vicinity of the SMBH \mbox{Sgr A$^{*}$} within roughly $1000\ r_G$, 
excluding the other possible VHE emitters in the
region (the pulsar-wind nebula candidate, the stellar cluster or the low-mass
X-ray binary system). Moreover, the position of a VHE emitting region could be
reconstructed with milli-arcsecond accuracy for the first time. A non-detection
could exclude the inner $\sim 1000 \ r_G$ around \mbox{Sgr A$^{*}$}  to be the dominant source of the
$\gamma$-emission from the direction of the Galactic Centre. 

The effect of PPEs can therefore
be used to probe the location of 
 particle acceleration processes in the environment of a quiescent super-massive
black hole. 
\section*{Acknowledgements}
AA and HSZ acknowledge the financial support from the BMBF under the contract number 05A08GU1. DH acknowledges the
International Space Science Institute (ISSI) in Bern for supporting a project related to the present research. We thank the anonymous referee for useful comments.

\bibliographystyle{mn2e}
\bibliography{abramowski_arxiv}

\appendix

\section{Re-scattered photons from inverse-Compton cascades}
\label{AppA}
In the following we calculate the re-scattered flux 
$f_\nu^{\rm casc}$, which is observed in addition to the attenuated $\gamma$-ray 
flux from \mbox{Sgr A$^{*}$}: 
\begin{equation} \label{eq:flux}
f_\nu^{\rm obs}  =  f_\nu \exp(-\tau) + f_\nu^{\rm casc},
\end{equation}
where $f_\nu$ labels the initial flux from \mbox{Sgr A$^{*}$}.
Since we are mainly concerned with the effect of the re-scattered photons
on the pair-production eclipse, the calculation will only consider the
first generation of cascade photons produced after pair-production
and inverse Compton up-scattering. While the cascade may continue, the subsequent scatterings will shift the mean energy of emerging photons well below 
the energies where the pair-eclipse can be observed until eventually the
energy threshold for pair production is not surpassed any more. The remaining
electrons and positrons will continue to cool. 
For reasons of simplified notation, we use in the
following energy in units of
rest mass of electron. Adapting the notation in the main body
of the text for the high energy photon $E = \epsilon_1 m_e c^2$, 
for the soft (stellar) photon $\varepsilon = \epsilon_2 m_e c^2$,
and for the electron/positrons $E_e = \gamma m_e c^2$.

\begin{figure}
\begin{center}
 \includegraphics[width=0.9\linewidth]{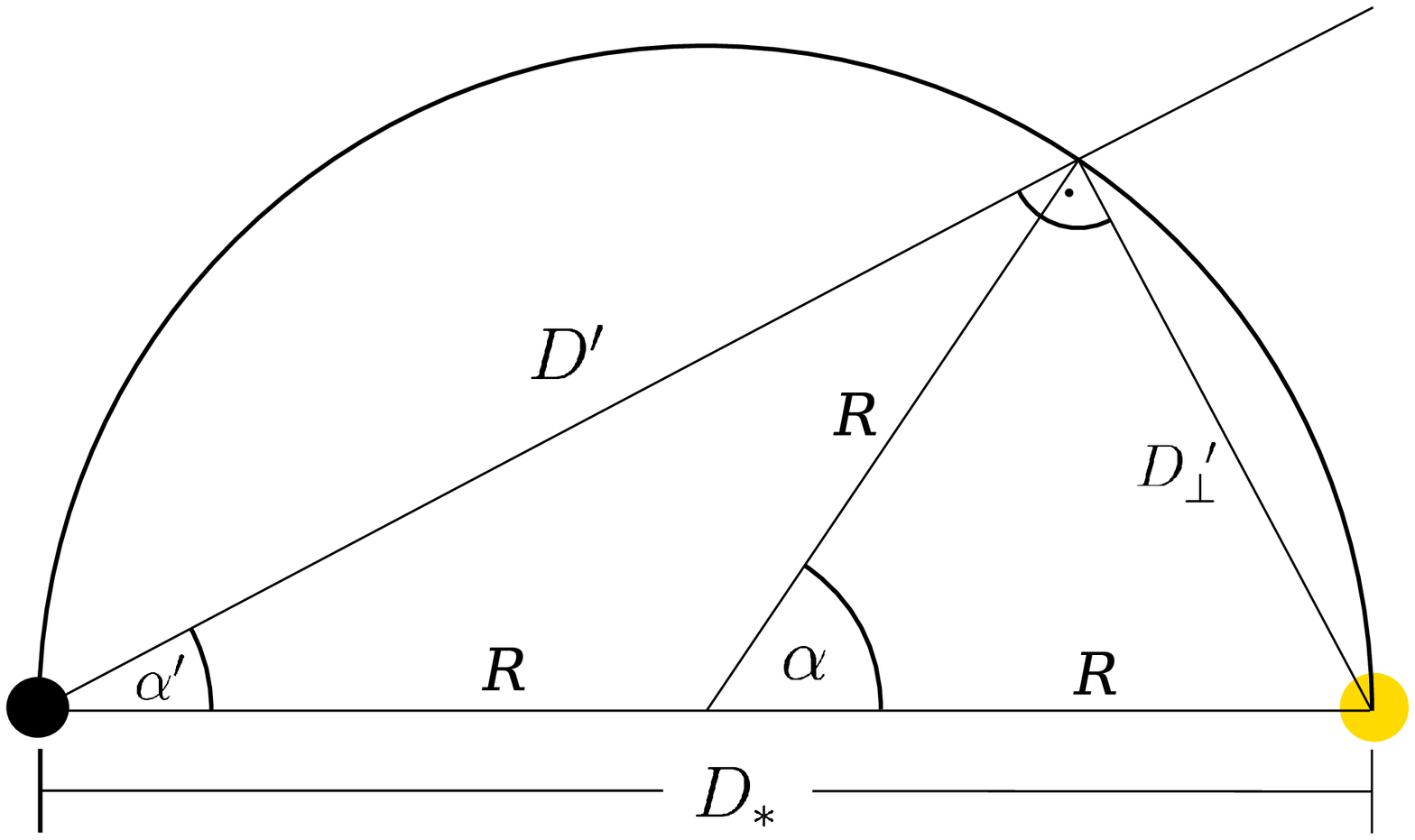}
 \caption{Geometry used for the calculation of the observed flux of re-scattered photons.}
\label{fig:intalpha}
\end{center}
\end{figure}

Given the fact, that the mean free path length for inverse Compton scattering
\mbox{($\mathcal{O}(>\,10^{16})$ cm)} is considerably larger than the gyro-radius in 
$\mu {\rm G}$ magnetic fields \mbox{($\mathcal{O}(<10^{14})$ cm)},
the electron/positron pair is quickly isotropised, so that the cascade
acts like an optically thin diffusor and wave-length shifter, replacing
the incoming photon of energy $\epsilon_1$ by two isotropically emitted
photons of lower energy. One of the two re-emitted photons produced
by inverse Compton scattering of the leading electron/positron carries a 
substantial fraction of the energy of the incoming photon. 
We choose to model the radiation transport by assuming that the $\gamma$-rays
are emitted isotropically from the $\gamma$-ray photosphere 
around the SMBH, which we 
define to be the region where $\tau\approx 1$ corresponding to the surface of 
a sphere of radius $\approx 20 \ r_G$. During the eclipse a fraction $1-\exp[-\tau(\epsilon_1)]$ of photons of
energy $\epsilon_1$ is absorbed along the line of sight subtending an
angle $\alpha'$ with respect to the line connecting the SMBH and the star. 
The pairs produced are then isotropised before upscattering photons from the
star to lower energies $\epsilon_1'$. The effect of cascading is thus
a shift of energy from $\epsilon_1$ to $\epsilon_1'$. Quantitatively, we 
calculate the probability $P(\epsilon_1'|\epsilon_1)$ for a given 
photon of energy $\epsilon_1$ to be re-emitted as a photon of energy $\epsilon_1'$. The geometry of the system can be simplified by assuming that the
cascade takes place at the point where the line of sight is closest to the 
stellar surface. This simplification is justified as the probability for
pair production is highest where the photon density is largest. In order to estimate the fraction of the re-scattered flux we also assume the SMBH as a point source. The resulting re-scattered flux can therefore be written in the following way
(see Fig. \ref{fig:intalpha} for a sketch of the underlying geometry):
\begin{equation}
\begin{split}
 f_\nu^{\rm casc}(\epsilon_1') &= \int d\epsilon_1\, 
P(\epsilon_1^\prime|\epsilon_1) f_\nu(\epsilon_1) \\ &\times \int d\phi \int d\alpha \,  
R^2 \sin{\alpha} \  \frac{1-\exp\left[-\tau(\alpha',\epsilon_1)\right]}{4\pi D^{\prime 2}},
\end{split}
\end{equation}
where the integration over the sphere $R=D_*/2$ incorporates all isotropically re-scattered components of the emission from the SMBH as seen by the observer. To account for the energy shift the integral has to be convolved with $P(\epsilon_1'|\epsilon_1)$.
Integrating over $\phi$ and using the geometric relations $D'=D_* \cos\alpha'$ and $\alpha'=\alpha/2$ reveals 
\begin{equation}
\begin{split}
f_\nu^{\rm casc} (\epsilon_1') =& \frac{1}{2} \int d\epsilon_1 P(\epsilon_1'|\epsilon_1)
f_\nu(\epsilon_1) \\ \times& \int\limits_{\alpha_\mathrm{min}}^{\alpha_\mathrm{max}}
d\alpha' \tan \alpha'\left(1-\exp\left[-\tau(\alpha',\epsilon_1)\right]\right). 
\end{split}
\end{equation}
The limit $\alpha_\mathrm{min}$ is given by the radius of the star while $\alpha_\mathrm{max}$ is defined by the 
$\gamma$-photosphere of the SMBH. The geometric effect results in a factor of $\approx 5\times 10^{-3}$ for VHE photons of energy $200\ {\rm GeV}$.

In order to evaluate the re-distribution kernel $P(\epsilon_1'|\epsilon_1)$, we use the differential pair-production cross section
$d\sigma/d\gamma$ in the kinematically accessible range:
$\epsilon_1 (1-\sqrt{1-(\epsilon_1\epsilon_2)^{-1})}<2\gamma<
 \epsilon_1 (1+\sqrt{1-(\epsilon_1\epsilon_2)^{-1})}$. This 
cross section is conveniently approximated \citep{1983Ap.....19..187A}:
\begin{equation}
\begin{split}
\frac{d\sigma}{d\gamma}(\epsilon_1,\epsilon_2,\gamma) = &\frac{3\sigma_T}{32 \epsilon_1^3 \epsilon_2^2} 
\left[\frac{4\epsilon_1^2}{\gamma (\epsilon_1-\gamma)}
\ln\left(\frac{4 \epsilon_2\gamma (\epsilon_1-\gamma)}{\epsilon_1}\right) - \right. \\
   8\epsilon_1\epsilon_2 + & \left.
 \frac{2\epsilon_1^2(2 \epsilon_1 \epsilon_2 - 1)}{\gamma (\epsilon_1-\gamma)} -
\left(1 - \frac{1}{\epsilon_1 \epsilon_2}\right) \frac{\epsilon_1^4}{\gamma^2 (\epsilon_1 - \gamma)^2}
\right].
\end{split}
\end{equation}
This cross-section is accurate to the level of a few per cent in the
limiting case of $\epsilon_1\gg \epsilon_2$
(see also discussion by \citet{1997A&A...324..395B}).
 The assumption of isotropy underlying this approximation
is not entirely justified. However, when evaluating the total 
cross section and the corresponding optical depth for the chosen geometry,
the relative differences to the line of sight integration of the
angle dependent cross section is a few per cent only. In order to
evaluate $P$, the differential cross section is normalised such that
\begin{equation}
 \begin{split}
 \frac{dn_{e}}{d\gamma}(\epsilon_1,\epsilon_2,\gamma) &= \frac{d\sigma}{d\gamma}(\epsilon_1,\epsilon_2,\gamma) \\
&\times  \left(0.5 \int d\epsilon\, n(\epsilon) \int d\gamma' \frac{d\sigma}{d\gamma'}(\epsilon_1,\epsilon,\gamma')\right)^{-1},
\end{split}
\end{equation}
taking into account that a pair of particles is produced. The
inverse Compton scattering cross section has been taken from \citet{1970RvMP...42..237B}
\begin{equation}
 \begin{split}
\frac{d\sigma}{d\epsilon_1'}(\epsilon_1',\epsilon_2,\gamma) &=
\frac{3\sigma_T}{4\epsilon_2 \gamma^2} 
\left[ 1 + \frac{z^2}{2(1-z)} + \frac{z}{b(1-z)} - \right.\\ 
    \frac{2z^2}{b^2(1-z)^2} &+ 
\left.    \frac{z^3}{2b(1-z)^2} - 
    \frac{2z \ln(b(1-z)/z)}{b(1-z)}
\right],
\end{split}
\end{equation}
with $z=\epsilon_1^\prime / \gamma$ and $b=4\epsilon_2\gamma$. Again, this differential
cross section is normalised,
\begin{equation}
\begin{split}
 \frac{dn}{d\epsilon_1'}(\epsilon_1',\epsilon_2,\gamma) &= 
 \frac{d\sigma}{d\epsilon_1'}(\epsilon_1',\epsilon_2,\gamma) \\
&\times\left( \int d\epsilon\,n(\epsilon) \int d\epsilon_1' \frac{d\sigma}{d\epsilon_1'}
(\epsilon_1',\epsilon,\gamma)
\right)^{-1},
\end{split}
\end{equation}
 so that finally
\begin{equation}
%\begin{split}
 P(\epsilon_1'|\epsilon_1) = \int d\gamma \int d\epsilon \, n(\epsilon) \frac{dn}{d\epsilon_1'}(\epsilon_1',\epsilon,\gamma) 
\int d\epsilon'\, n(\epsilon') \frac{dn_e}{d\gamma}(\epsilon_1,\epsilon',\gamma).
%\end{split}
\end{equation}
\begin{figure}
 \includegraphics[width=0.9\linewidth]{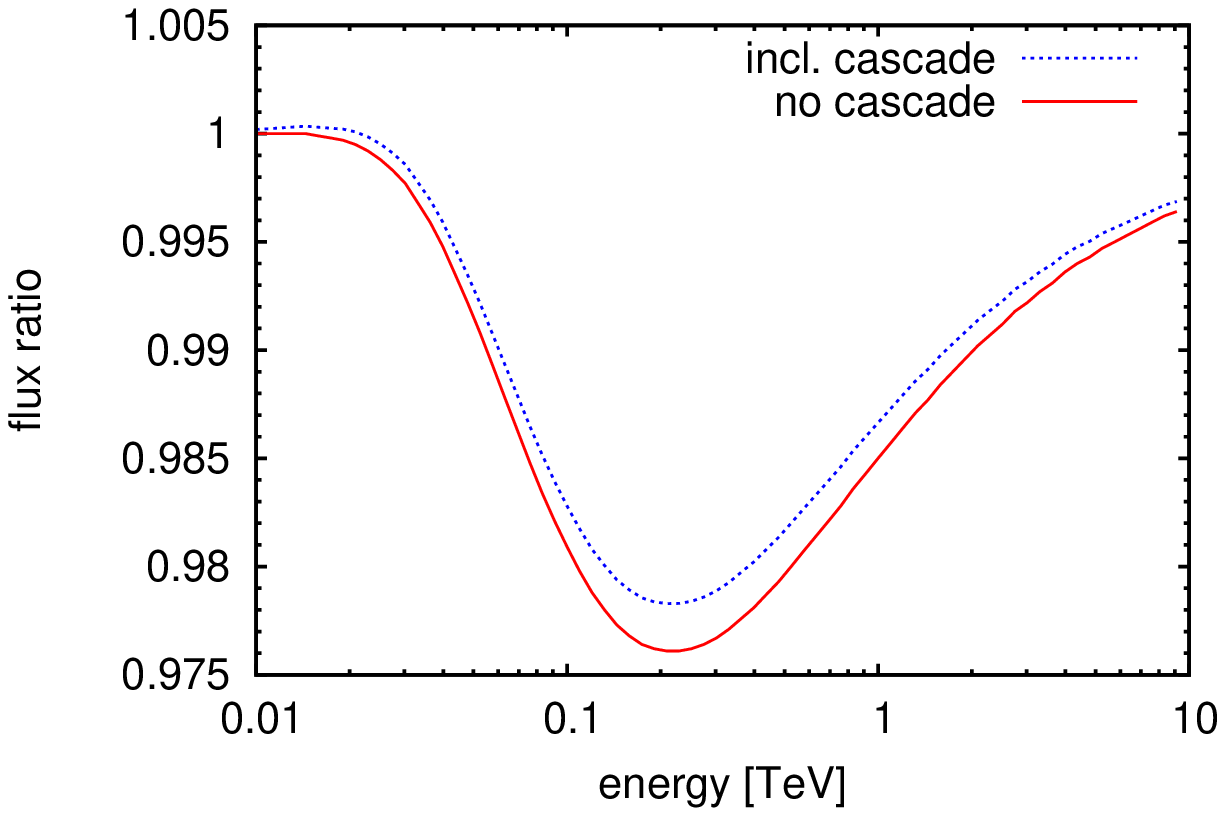}
 \caption{Flux ratio $f^\mathrm{obs}_\nu/f_\nu$ for the star S14 at the time of the largest absorption effect with (dashed blue) and without (solid red) the contribution of the photons from the cascade.}
\label{fig:result}
\end{figure}

In Fig. \ref{fig:result} we display the effect of the cascade by showing
the ratio of the observed flux to the initial flux with and without
the additional cascade component. 
The result indicates a small increase of the flux when including the
first generation cascade. The resulting change is similar to the uncertainties
resulting from the uncertainties of the orbital elements. The slight bump due to the isotropic
cascade radiation could in principle be observed even during the passage of a
star far away from the line of sight (thus leading to no absorption effect).
The effect is however at the level of 0.1 per cent and therefore not easily 
observable.

The additional feature (pile-up) in the energy spectrum at
energies below the main absorption peak has a
timing signature shifted with respect to the 
time-dependent absorption due to the cooling time of the electrons, which is in $\mathcal{O}(20)\ {\rm d}$. 
The resulting (time-dependent) 
emission spectrum of electrons cooling in the Klein-Nishina
regime could in principle be calculated using the approach suggested by \citet{2007A&A...474..689M}. Given the smallness of the effect however, we consider it of little interest here.

\section{Parametrisation for the effective areas}
\label{AppB}
For the effective areas $A_{\mathrm{eff}}(E)$ used in the simulation of the light-curves (Eq. \ref{eq:rate}) we used the following parametrisation:

\begin{equation}
  A_{\mathrm{eff}}(E) = g_1 \left(\frac{E}{\mathrm{MeV}}\right)^{-g_2} \exp \left(-\frac{g_3}{E}\right).
    \label{eq:aeff}
\end{equation}
The parameters for the three different IACTs are given in \mbox{Table \ref{tab:para}}. For H.E.S.S. and H.E.S.S. II the curve is based upon the effective area from \citet{2006A&A...457..899A} and \citet{Punch2005} respectively. For CTA we assume the effective area to increase by a factor of 25 as given in \citet{2008ICRC....3.1313H}.
\begin{table}
\caption{ \label{tab:para} Parameters used for the calculation of the effective area for the three different IACTs.}
\begin{center}
\begin{tabular}{|l|c|c|c|}
\hline
&H.E.S.S.&H.E.S.S. II&CTA\\
\hline
$g_1$ & $6.85\times 10^{9}\, \mathrm{cm}^2$ & $2.05\times 10^{10}\, \mathrm{cm}^2$ & $1.71\times 10^{11}\, \mathrm{cm}^2$  \\
$g_2$ & 0.0891  & 0.0891 & 0.0891  \\
$g_3$ & $5\times 10^{5} \,\mathrm{MeV}$  & $1\times 10^{5} \,\mathrm{MeV}$ & $1\times 10^{5} \,\mathrm{MeV}$  \\
\hline
\end{tabular}
\end{center}
\end{table}

\end{document}